\documentstyle[12pt]{article}
\begin{document}
\title{The matrix element for radiative Bhabha scattering
       in the forward direction}
\author{Ronald Kleiss\\ NIKHEF-H\\ Amsterdam, The Netherlands}
\date{}
\maketitle
\newcommand{\bq}{\begin{equation}}
\newcommand{\eq}{\end{equation}}
\newcommand{\bqa}{\begin{eqnarray}}
\newcommand{\eqa}{\end{eqnarray}}
\newcommand{\nl}{\nonumber\\}
\newcommand{\eqn}[1]{Eq.(\ref{#1})}
\newcommand{\umu}{^{\mu}}
\newcommand{\unu}{^{\nu}}
\newcommand{\lmu}{_{\mu}}
\newcommand{\de}{\Delta}
\newcommand{\mat}{\mid\!\begin{cal}M\end{cal}\!\mid^2}
\newcommand{\ormu}[1]{\begin{cal}O\end{cal}\left(\mu^{#1}\right)}
\newcommand{\bc}{\mbox{\begin{bf}C\end{bf}}}
\newcommand{\mdot}{\!\cdot\!}

\begin{abstract}
We present an approximation to the matrix
element for the process $e^+e^-\to e^+e^-\gamma$, appropriate to
the situation where one or both of the fermions
are scattered over very small angles.
The leading terms in the situation where all scattering angles are small
contains not only terms quadratic in the electron mass, but also
quartic and even sextic terms must be included.
Special attention is devoted to
the numerical stability of the resultant expression. Its relation to
several existing formulae is discussed.
\end{abstract}
For any $e^+e^-$ collider, the process of Bhabha scattering is very important.
Whereas large-angle scattering provides its share of information on
QED or theories of electroweak interactions, small-angle scattering serves
as a tool for accurate luminosity determination; in each of these cases,
radiative corrections (and hence bremsstrahlung) cannot be neglected. But the
radiative process by itself, namely
\bq
e^+(p_1)e^-(q_1)\to e^+(p_2)e^-(q_2)\gamma(k)\;\;,
\label{theprocess}
\eq
is important in its own right, especially when one or either of the
electrons is scattered over a small, or even negligible, angle.
For instance, such events form an important background to low-energy
photon-photon processes, as recently pointed out,
for the DA$\Phi$NE collider, in \cite{greco};
moreover, interestingly, the process (\ref{theprocess}) is, for machines
such as LEP, the most important restriction on the beam lifetimes
(see, for instance, \cite{hbu}). Since the Bhabha cross section peaks
at small fermion scattering angles, and the photon is preferentially
emitted more or less collinearly with a charged particle, such processes
are dominated by events where all scattering angles are small, and
many different momentum transfers become small simultaneously.
It is therefore important to have reliable expressions for the matrix
element (we shall always understood it to be squared and summed/averaged
over all particle spins) in these extreme kinematical configurations.
An allowed simplification of the problem,
if we restrict ourselves to small fermion scattering angles,
is to neglect the $s$-channel diagrams,
and the interference between radiation from the
positron and that from the electron line. Hence, in what follows we shall
assume that we only have the two Feynman diagrams in which
the photon is emitted by the incoming and outgoing positron (radiation
from the electron is then trivially obtained by $e^+\leftrightarrow e^-$).
We introduce the following notation:
\bqa
P\umu & = & p_1\umu+q_1\umu = p_2\umu+q_2\umu+k\umu\;\;,\nl
q\umu & = & q_1\umu-q_2\umu\;\;,\nl
s & = & P\mdot P \equiv 4E^2\;\;,\nl
t & = & q\mdot q\;\;,\nl
\de_{1,2} & = & p_{1,2}\mdot k\;\;,\nl
s_1 & = & (p_2+q_2)^2 = (P-k)^2\;\;.
\eqa
The five invariants $s$, $s_1$, $\de_{1,2}$,
and $t$, specify the kinematics completely up to a trivial azimuthal
rotation around the beam axis.
Using the algebraic program {\tt FORM} \cite{form}, we then
find for the matrix element, $\mat$:
\bqa
{1\over e^6}\mat & = &
{1\over\de_1\de_2(-t)}\left[
 (s-6m^2)^2+(s+4m^2-2\de_2+t)^2\right.\nl
& & \hphantom{{1\over\de_1\de_2(-t)}}
+(s_1-6m^2)^2+(s_1+4m^2+2\de_1+t)^2\nl
& & \hphantom{{1\over\de_1\de_2(-t)}}\vphantom{{1\over\de_1}}\left.
 \vphantom{(s)^2}-12m^2(t+10m^2)\right]\nl
& & +{8m^2\over\de_1\de_2 t^2}\left[
 (s-2m^2-2\de_2+t)(s_1-2m^2-2\de_1+t)\right.\nl
& & \hphantom{{8m^2\over\de_1\de_2 t^2}}\left.
 -(\de_1-\de_2)^2\right]\nl
& & -{2m^2\over\de_1^2t^2}\left[
 (s_1-2m^2)^2+(s_1-2m^2+t)^2+4m^2t\right]\nl
& & -{2m^2\over\de_2^2t^2}\left[
 (s-2m^2)^2+(s-2m^2+t)^2+4m^2t\right]\;\;,
\label{fullexpression}
\eqa
where $m$ is the electron mass. If we request the bremsstrahlung photon to
have a non-infinitesimal energy, there is only one smallness
parameter in our problem, namely $\mu=m/E$; we shall now discuss
what is the order of $\mu$ of the various invariants in the
different collinear situations. To this end, note that with a relative
error of $\ormu{2}$ we can write
\bq
\mid\!t\!\mid = {m^2\de_2^2\over sEE'}
+EE'\left(1-\cos<(\vec{q}_1,\vec{q}_2)\right)\;\;,
\eq
where $E'=E(1-2\de_2/s)$ is the energy of $q_2$.
We can therefore distinguish the following collinear situations,
in which the cross section displays strong peaks:
\begin{itemize}
\item $\bc_1$: $\de_1=\ormu{2}$, $\de_2,t=\ormu{0}\;\;$;
\item $\bc_2$: $\de_2=\ormu{2}$, $\de_1,t=\ormu{0}\;\;$;
\item $\bc_3$: $t=\ormu{2}$, $\de_{1,2}=\ormu{0}\;\;$;
\item $\bc_4$: $\de_{1,2}=\ormu{2}$, $t=\ormu{6}\;\;$.
\end{itemize}
Unless $k^0$ is very close to $E$, both $s$ and $s_1$ are $\ormu{0}$.
Allowed approximate matrix elements are those in which only the
leading orders in $\mu^{-1}$ are kept.

We can now relate \eqn{fullexpression} to the old CALKUL result
of \cite{calkul} by
noting that in $\bc_1$ and $\bc_2$ we may neglect $m^2$ with respect to
$s$, $s_1$, and $t$, but not to $\de_{1,2}$, so we find
\bqa
{1\over e^6}\mat & \rightarrow &
{s^2+s_1^2+z^2+z_1^2\over\de_1\de_2(-t)}\nl
& & -{m^2\over\de_1^2}{2(s_1^2+z_1^2)\over t^2}
      -{m^2\over\de_2^2}{2(s^2+z^2)\over t^2}\;\;,\nl
z & = & 2\de_2-s-t = (p_1-q_2)^2+\ormu{2}\;\;,\nl
z_1 & = & -2\de_1-s_1-t = (p_2-q_1)^2+\ormu{2}\;\;,
\eqa
which is precisely the CALKUL result
for these two diagrams. Obviously,
this expression is not appropriate to either $\bc_3$ or $\bc_4$.
We therefore proceed to drop, in \eqn{fullexpression}, terms that
are of the order of $m^2/\de_1\de_2t$, which are always negligible,
and we find
\bqa
{1\over e^6}\mat & = & A_0+A_1+A_2+A_3+A_4+A_5+A_6+A_7+A_8\;\;,\nl
A_0 & = & \left(s^2+s_1^2+z^2+z_1^2\right)/(-\de_1\de_2t)\;\;,\nl
A_1 & = & -4m^2Z^2/(\de_1^2\de_2^2t^2)\;\;,\nl
A_2 & = & -8m^2(\de_1^2+\de_2^2)/(\de_1\de_2t^2)\;\;,\nl
A_3 & = & 16m^4Z(\de_1-\de_2)/(\de_1^2\de_2^2t^2)\;\;\nl
A_4 & = & -16m^6(\de_1-\de_2)^2/(\de_1^2\de_2^2t^2)\;\;\nl
A_5 & = & -4m^2s_1/(\de_1^2t)\;\;,\nl
A_6 & = & -4m^2s/(\de_2^2t)\;\;,\nl
A_7 & = & -2m^2/\de_1^2\;\;,\nl
A_8 & = & -2m^2/\de_2^2\;\;,\nl
Z & = & s\de_1-s_1\de_2-2\de_1\de_2\;\;.
\label{dropmasses}
\eqa
Let us tabulate the formal power of $\mu$ of each term in the four
distinct collinear situations, writing $\zeta$ for the formal power of
$Z$ in $\bc_4$:
\begin{center}\begin{tabular}{|c||c|c|c||c|c|c|c|c|c|c|c|c|}
\hline\hline
 & $\de_1$ & $\de_2$ & $t$ & $A_0$ & $A_1$ & $A_2$ &
 $A_3$ & $A_4$ & $A_5$ & $A_6$ & $A_7$ & $A_8$ \\
\hline
$\bc_1$ & 2 & 0 & 0 & -2 & -2 & 0 & 0 & +2 & -2 & +2 & -2 & +2 \\
$\bc_2$ & 0 & 2 & 0 & -2 & -2 & 0 & 0 & +2 & +2 & -2 & +2 & -2 \\
$\bc_3$ & 0 & 0 & 2 & -2 & -2 & -2 & 0 & +2 & 0 & 0 & +2 & +2 \\
$\bc_4$ & 2 & 2 & 6 & -10 & 2$\zeta$-18 & -10 &
 $\zeta$-14 & -10 & -8 & -8 & -2 & -2 \\
\hline\hline
\end{tabular}\end{center}
We note the following points. In order to avoid negative $\mat$, we need
to have $\zeta\ge4$ in $\bc_4$: we shall see that this is indeed the case.
Furthermore, the terms $A_{5,6,7,8}$ are always negligible when $t$
is small. This is a reasonable approximation once we have assumed that
the interference between radiation from the $e^+$ and the $e^-$ can be
discarded. Lastly, we have to ensure that the form (\ref{dropmasses})
can be evaluated in a numerically stable manner. To this end we must
control the cancellations inside $Z$, and in addition prove that the
matrix element $\mat$ is indeed $\ormu{-10}$ in $\bc_4$. These two last
items will be solved simultaneously at the end of our discussion.
Summarizing the above considerations, we propose the following form for
the matrix element, valid for small $t$:
\bqa
{1\over e^6}\mat & = &
   {1\over-\de_1\de_2t}
   \left(s^2+s_1^2+(s+t-2\de_2)^2+(s_1+t+2\de_1)^2\right)\nl
& & -{4m^2\over\de_1^2\de_2^2t^2}
    (s\de_1-s_1\de_2-2\de_1\de_2)^2
    -{8m^2\over\de_1\de_2t^2}(\de_1^2+\de_2^2)\nl
& & +{16m^4\over\de_1^2\de_2^2t^2}
    (\de_1-\de_2)(s\de_1-s_1\de_2-2\de_1\de_2)\nl
& & -{16m^6\over\de_1^2\de_2^2t^2}(\de_1-\de_2)^2\;\;.
\label{final}
\eqa

Before proceeding, let us compare the expression of \eqn{final}
with existing ones.
An earlier form of the matrix element, due to Baier {\it et al.\/}
\cite{baier} and Altarelli and coworkers \cite{altarelli}, reads
\bq
{1\over e^6}\mat = {2(s^2+s_1^2)\over-\de_1\de_2t}
-{4m^2\over\de_1^2\de_2^2t^2}(s\de_1-s_1\de_2)^2
-{8m^2\over\de_1\de_2t^2}(\de_1^2+\de_2^2)\;\;.
\label{altarel}
\eq
We see that \eqn{altarel} differs from \eqn{final} in that it has
a different form for $Z$, and no terms with $m^4$ or $m^6$ ---
which, however,
cannot be neglected, in contrast to what is commonly the case with
ultrarelativistic matrix elements.

An alternative approximation to the matrix element was recently
proposed in \cite{greco}.  It includes terms that have an explicit
$m^2$, in the following combination, for which we use the notation of
the present paper:
\bqa
G & \equiv &
  -{4m^2\over\de_1^2}\left({s_1\over t}+{t\over s_1}+1\right)^2
  -{4m^2\over\de_2^2}\left({s\over t}+{t\over s}+1\right)^2\nl
& &
  -{4m^2\over\de_1\de_2}\left(
  {2s(p_2-q_1)^2+3(p_1-p_2)^4\over t^2}\right.\nl
& & \hphantom{-{4m^2\over\de_1\de_2}(}\left.+
  {s+(p_2-q_1)^2-2(p_1-p_2)^2\over t}+1\right)\;\;.
\eqa
Let us first note that this expression is not invariant
under time-reversal, which in our problem is the operation
$p_1\umu\leftrightarrow p_2\umu$,
$q_1\umu\leftrightarrow q_2\umu$, and $k\umu\leftrightarrow-k\umu$.
What may be even more serious is the peaking behaviour. Expressing
$G$ in terms of the five independent invariants $s$, $s_1$, $t$ and
$\de_{1,2}$, and retaining only those terms that contribute in
$\bc_3$ or $\bc_4$, we have
\bq
G \sim -{4m^2(s\de_1-s_1\de_2)^2\over\de_1^2\de_2^2t^2}
           -{48m^2(\de_1-\de_2)^2\over\de_1\de_2t}
           +{16m^2s(\de_1-2m^2)\over\de_1\de_2t^2}\;\;.
\eq
The first two of these terms are $\ormu{-10}$ in $\bc_4$,
but the last is $\ormu{-12}$, since in general the combination
$\de_1-2m^2$ is $\ormu{2}$ rather than $\ormu{4}$. We therefore
have to conclude that, at least in $\bc_4$, the result of
\cite{greco} is not appropriate. Note, however, that in the
soft-photon limit disucssed in \cite{greco}, $\de_1\to0$ so
that the correct $\ormu{10}$ behaviour is recovered in that case.\\

Let us now turn to the numerical stability of the expression
(\ref{final}). We may assume that we can evaluate $\de_{1,2}$ and $t$
in a stable manner for given momenta. It remains to control the
cancellation from $\ormu{2}$ down to $\ormu{4}$ inside $Z$,
that occurs in $\bc_4$. To this end we introduce $y=q^0/E$, $u=1-y$, and
a vector $r\umu$, by $q_2\umu = uq_1\umu +r\umu$. Denoting by the subscript
$L$ components along $\vec{q}_1$, and by $T$ the orthogonal components,
we find
\bqa
r\umu& = & (r^0,\vec{r}_T,r_L)\;\;,\nl
r^0 & = & 0\;\;,\nl
\vec{r}_T & = & -\vec{q}_T\;\;,\nl
r_L & = & (-2m^2y+t)/(2p)\;\;,\nl
r\mdot r & = & m^2y^2+ut\;\;,
\eqa
with $p=\mid\!\vec{p}_1\!\mid$.
In $\bc_4$, then, $\vec{r}_T$ is $\ormu{3}$, and $r_L$ is $\ormu{4}$,
since $y=\ormu{2}$.
Some straightforward algebra now leads to the surprisingly simple form
\bq
Z = s(k\mdot r)\;\;.
\eq
The best frame to evaluate this expression is the rest frame of
$R\umu \equiv p_2\umu+k\umu$.
Denoting the various components in this frame by carets, we have,
with $w=\sqrt{R\!\cdot\!R}=\ormu{2}$:
\bqa
\hat{r}^0 & = & -(2m^2y+t)/(2w)\;\;,\nl
\hat{\vec{r}}_T & = &
-\vec{q}_T\left(1+{\hat{r}^0\over R^0+w}\right)\;\;,\nl
\hat{r}_L & = & r_L+(p-q_L){\hat{r}^0\over R^0+w}\;\;,\nl
\hat{k}^0 & = & (w^2-m^2)/(2w)\;\;.
\eqa
we see that all components of $\hat{r}\umu$ are $\ormu{3}$, and
$\hat{k}\umu$ is $\ormu{}$. Since $r\mdot r$ is $\ormu{6}$, no further
cancellations are to be expected\footnote{If
$r\umu r\lmu$ was smaller, yet another cancellation
might be lurking in the product $k\mdot r$.},
and $Z$ is indeed $\ormu{4}$.\\

Finally, we prove that, in $\bc_4$, the matrix element is indeed
$\ormu{-10}$. In this matrix element, the electron line and
the virtual photon propagator enter via the
object
\bqa
H^{\mu\nu} & = &{1\over t^2}\sum\limits_{\mbox{spins}}
 \bar{v}(q_1)\gamma\umu u(q_2)\bar(q_2)\gamma\unu v(q_1)\nl
& & {4\over t^2}\left(q_1\umu q_2\unu + q_1\unu q_2\umu +
 {t\over2}g^{\mu\nu}\right)\;\;.
\eqa
The last term cancels one of the factors $t$, and will hence not
contribute in $\bc_4$. We now rewrite $q_{1,2}\umu$ in terms of $q\umu$
and $r\umu$:
\bq
q_1\umu=(r\umu+q\umu)/y\;\;\;,\;\;\;
q_2\umu=(r\umu+uq\umu)/y\;\;.
\eq
Owing to current conservation, the terms with $q\umu,q\unu$ evaluate to
zero, and the relevant part of the electron current is therefore
\bq
H^{\mu\nu} = {8r\umu r\unu\over y^2t^2}\;\;.
\eq
This object, when evaluated in the $R\umu$ rest frame, is itself of
order $\ormu{-10}$ as we have seen above. In addition, again in the
$R\umu$ rest frame, the positron current is $\ormu{0}$ since
there
\bqa
p_{1,2}\umu,k\umu & = & \ormu{}\;\;,\nl
\de_{1,2} & = & \ormu{2}\;\;,\nl
u(p_{1,2}) & = & \ormu{1/2}\;\;.
\eqa
This finishes the argument: $\mat$ is indeed $\ormu{-10}$ in $\bc_4$,
and we can evaluate it in a numerically stable manner, without the
need for extended precision ({\tt REAL*16}) as in \cite{greco} to
control the cancellations. An application of the form
(\ref{final}), to the study of beam-beam-strahlung as a limitation
to the LEP beam lifetimes, will be presented elsewhere \cite{helmut}.

\end{document}